\documentclass[12pt]{article}
\input epsfig.sty
\newsavebox{\sboxpubnumber}
\newsavebox{\sboxpubdate}

\newcommand{\pubnumber}[1]{\begin{lrbox}{\sboxpubnumber}
{\begin{tabular}{l} #1 \\
				 \usebox{\sboxpubdate}
				 \end{tabular}}
                           \end{lrbox}
                           \pubblock}
\newcommand{\Title}[1]{\begin{center} {\Large #1 } \end{center}}
\newcommand{\Author}[1]{\begin{center}{ \sc #1} \end{center}}
\newcommand{\Address}[1]{\begin{center}{ \it #1} \end{center}}

\newcommand{\pubblock}{\rightline{
			\usebox{\sboxpubnumber}}}
\newenvironment{Abstract}{\begin{quotation}  }{\end{quotation}}
\newenvironment{Presented}{\begin{quotation} \begin{center}
             PRESENTED AT\end{center}\bigskip
      \begin{center}\begin{large}}{\end{large}\end{center}
      \end{quotation}}

\def\vf{\phi}
\def\laq{\raise 0.4ex\hbox{$<$}\kern -0.8em\lower 0.62 ex\hbox{$\sim$}}
\def\gaq{\raise 0.4ex\hbox{$>$}\kern -0.7em\lower 0.62 ex\hbox{$\sim$}}

\begin{document}
\begin{titlepage}
\pubnumber{CERN-TH/2002-371} 
\vfill
{}\Title{MAGNETOGENESIS, VARIATION OF GAUGE COUPLINGS AND INFLATION}
\vskip 2 cm
\Author{Massimo Giovannini\footnote{Electronic address: 
 Massimo.Giovannini@cern.ch}}
\Address{
Theoretical Physics Division, CERN, CH-1211, Geneva, Switzerland}
\vskip 2cm
\begin{Abstract}
The interplay among the possible variation 
of gauge coupling and the inflationary dynamics is 
investigated in a simplified toy model.
Depending upon various parameters (scalar mass, curvature scale at the 
end of inflation and at the onset of the radiation epoch), 
the two-point function of 
the magnetic inhomogeneities grows during the de Sitter stage of expansion
and, consequently, large scale magnetic fields are generated. 
The requirements coming from inflationary magnetogenesis are examined 
together with the theoretical constraints stemming from the 
explicit model of the evolution of the gauge coupling.
\end{Abstract}
\vfill
\begin{Presented}
     9th ``Course of Astrofundamental Physics''\\
     ``International School D. Chalonge'''\\
   Palermo, September 2002.
\end{Presented}
\end{titlepage}
\renewcommand{\theequation}{1.\arabic{equation}}
\setcounter{equation}{0} 
\section{Introduction} 

Suppose  that during a de Sitter stage of expansion the 
coupling constant of an Abelian gauge field evolves in time. Thus 
the kinetic term of the 
gauge field can be written (in four space-time
dimensions) as 
\begin{equation}
S_{\rm em} = -\frac{1}{4}\int d^4 \, x \sqrt{- G} f(\phi) 
F_{\alpha\beta} F^{\alpha\beta},
\label{ac1}
\end{equation}
where $G$ is the determinant 
of the space-time metric, $\phi$ is a scalar field
(which can depend upon space and time) and $g(\phi)= f(\phi)^{-1/2}$ 
is the coupling \footnote{The Heaviside 
electromagnetic system of units will be used throughout the investigation. 
The effective ``electron'' charge will then be given, in the 
present context, by $ e(\phi) = e_1 f(\phi)^{-1/2}$.}. 
This type of vertex is typical of scalar-tensor theories 
of gravity \cite{bg} 
and of the low energy string effective action \cite{mmv}. Early suggestions
that the Abelian gauge coupling may change over 
cosmological times were originally made by Dirac \cite{dir1}
(and subsequently discussed in \cite{dir2} and in \cite{dir3}) 
mainly in the framework 
of ordinary electromagnetism.

The dynamics of the field $\phi$ will 
be described by the action of a minimally coupled 
(massive) scalar. If the field is displaced from the minimum 
of its potential during inflation, there will be a phase where 
 the field relaxes. Provided 
the scalar mass is much smaller than the curvature scale during inflation
such a phase could be rather long.
During the de Sitter stage, 
the specific form of the expanding background will dictate, through
the equations of motion, the rate of suppression of the amplitude of $\phi$. 

While the field relaxes toward the minimum of its potential, 
energy is pumped from the  homogeneous mode of $\phi$ to the 
gauge field fluctuations. 
The function $f(\phi)$ can be either an increasing 
function of $\phi$ (leading to a decreasing coupling) 
or a decreasing function of $\phi$ (leading to an increasing coupling).
In both cases, depending upon the parameters of the 
model,  the two-point 
correlation function of magnetic inhomogeneities 
increases during the inflationary stage. This implies that 
large scale magnetic fields can be potentially generated.

The implications of large scale magnetic fields 
for cosmology, astrophysics and high-energy physics 
have been recently reviewed in \cite{ch}  where 
various useful references are reported. The interested student will 
find in \cite{ch} a morte physical introduction to the subject of large 
scale magnetic fields. The toy model 
discussed in the present paper is only 
an useful exercise suggesting a possible 
 connection between the variation of gauge couplings and the 
production of large scale magnetic fields. 

Recently \cite{webb} the variation in the fine structure 
constant has been suggested by observations of absorption lines 
at different frequencies. The results of these claims can be summarized by saying that
$\alpha_{\rm em}$ was smaller in the past. 
Defining, indeed $\Delta \alpha_{\rm em} = \alpha_{\rm today} - \alpha_{\rm past}$,
measurements suggest that
\begin{equation}
\frac{\Delta \alpha_{\rm em} }{\alpha_{\rm em} } = (0.72 \pm 0.18) 10^{-5}.
\end{equation}
To avoid confusions, in the present scenario, gauge couplings 
are {\em constant} today and their variation 
occurs prior to big-bang nucleosynthesis (BBN).

The gauge coupling has to be (almost)
constant by the time when the Universe is 
approximately old of one second, namely by the 
time of BBN. 
The abundances of light elements are very sensitive 
to any departure from the standard cosmological model.
Hence, fluctuations in the baryon to photon ratio,
matter--antimatter domains, 
anisotropies in the expansion of the four space-time dimensions, 
can all be successfully constrained by demanding that the abundances 
of the light elements are correctly reproduced. Following the same logic 
the variation in the gauge couplings can also be constrained from BBN \cite{coup}.

The plan of the present contribution is then the following. 
In Section II the basic ideas concerning the 
model of evolution of the gauge coupling 
will be introduced. In Section III 
bounds coming both from the homogeneous and from the 
inhomogeneous evolution of $\vf$ will 
be described. In Section IV the evolution of the 
magnetic inhomogeneities will be addressed along the  various 
stages of the model with particular attention to the 
role of the two-point function. 
In Section V the large scale magnetic fields produced 
in the scenario will be estimated. Section VI contains 
some concluding remarks.
The considerations reported in the present paper extend 
and complement the results of \cite{mg}. 

Finally, in concluding this Introduction, I wish to 
remind a couple of useful references of two other 
lecturers of the Chalonge school. In
Ref. \cite{raf}  an interesting discussion 
on the possible ambiguities arising in the observations of large scale
fields in clusters is reported. In \cite{danhector} another 
idea on the generation of magnetic fields is reported.

\renewcommand{\theequation}{2.\arabic{equation}}
\setcounter{equation}{0}
\section{Basic Equations}  
Thanks to the high degree of isotropy and homogeneity 
of the observed Universe, the background geometry 
can be  described  using a  (conformally flat) 
Friedmann-Robertson-Walker (FRW) line element
\begin{equation}
ds^2 = G_{\mu\nu} dx^{\mu} dx^{\nu} =
 a^2(\eta)[ d\eta^2 - d \vec{x}^2],
\label{metric}
\end{equation}
where $\eta$ is the conformal time coordinate and $G_{\mu\nu}$ is the 
four-dimensional space-time metric. The cosmic time coordinate 
(often employed in this investigation) is related to $\eta$ as 
$a(\eta)~d\eta = d t$.

From the anisotropies of the Cosmic Microwave Background (CMB) 
it is consistent to assume that the Universe underwent 
a period of inflationary expansion of de Sitter or quasi-de Sitter type. 
Therefore $a(\eta) \sim -\eta_{1}/\eta$ during a phase stopping, approximately,
when the curvature scale was $H_{1} \leq 10^{-6} M_{\rm P}$. For  $\eta > 
\eta_1$ (possibly after a transient period) the Universe  gets 
dominated by radiation [i.e. $a(\eta) \sim \eta$] and then, after 
decoupling, by dust matter [i.e. $a(\eta) \sim \eta^2$].

The action describing the  dynamics of the (Abelian) gauge coupling in a given 
background geometry can be  parametrized as
\begin{equation}
S= \int d^4 x \,\,\sqrt{-G} \biggl[ \frac{1}{2} G^{\mu\nu} \partial_{\mu}
 \vf \partial_{\nu} \vf 
- V(\vf) - \frac{1}{4 g^2(\vf)} F_{\mu\nu} F^{\mu\nu}
\biggr].
\label{action1}
\end{equation}
In Eq. (\ref{action1}) $\phi$ {\em is not the inflaton} 
but an extra (scalar) degree of freedom taking care of the 
evolution of the gauge coupling.

From Eq. (\ref{action1}) the equations of motion can be derived
\begin{eqnarray}
&& \frac{1}{\sqrt{-G}} \partial_{\mu}\biggl[ \sqrt{- G} G^{\mu\nu} 
\partial_{\nu} \vf \biggr]
+ \frac{\partial V}{\partial \vf} = 
\frac{1} { 2 g^3(\vf) } \frac{\partial g}{\partial \phi} 
F_{\alpha \beta} F^{\alpha\beta}, 
\label{Eq1}\\
&& \frac{1}{\sqrt{- G}} \partial_{\alpha} \biggl[ 
\frac{\sqrt{- G}}{g^2(\vf)} F^{\alpha\beta} \biggr]
= 0.
\label{Eq2}
\end{eqnarray}

If the gauge coupling does not change, 
the evolution of Abelian gauge fields is conformally  invariant. 
Hence,  using the the conformal time coordinate, 
the appropriately rescaled gauge field amplitudes obey 
a set of equations which is exactly the one they 
would obey Minkowski space. 
In order to simplify the explicit form of the equations of motion 
the electric and magnetic fields will be 
rescaled in such a way that the obtained system of equations 
will reproduce the usual (conformally invariant) system in the 
limit $g\rightarrow $ constant. The rescalings in the fields are 
\begin{eqnarray}
&&{\vec{B}}=a^2 \vec{{\cal B }}, \,\,\,\vec{E}=a^2\vec{{\cal E}},~~~~
\vec{A}= a \vec{\cal A},
\label{resc}
\end{eqnarray}
where  $\vec{{\cal
B}}$, $\vec{{\cal E}}$, $\vec{{\cal A}}$,
 are  the flat-space quantities whereas
$\vec{B}$, $\vec{E}$, $\vec{A}$,  are the
curved-space ones.

The explicit form of  our system  becomes, in the metric (\ref{metric}):
\begin{eqnarray}
&& \frac{\partial^2\vf}{\partial\eta^2} 
+  2 {\cal H} \frac{\partial \vf}{\partial\eta}  
+ \frac{\partial V}{\partial \phi} a^2 - \nabla^2 \phi 
=  \frac{ 1}{ 2 g^3 a^2 }
\frac{\partial g}{\partial \phi} \biggl[ \vec{E}^2 - \vec{B}^2 \biggr] ,
\label{as1}\\
&& \frac{\partial{\vec{B}}}{\partial\eta} = -\vec{\nabla}
\times \vec{{E}}, 
\label{as1b}\\
&&\frac{\partial}{\partial\eta}\biggl[ \frac{1}{g^2(\vf)} \vec{E}\biggr] 
 =
\frac{1}{g^2(\vf)} \biggl[{\vec{\nabla}}\times \vec{B} - \frac{2}{g} 
\frac{\partial g}{\partial\phi} \vec{\nabla}\phi\times \vec{B}\biggr], 
\label{as2}\\
&& {\vec{\nabla}}\cdot{\vec{B}}=0,
\label{as2b}\\
&&{\vec{\nabla}}\cdot {\vec{E}}= \frac{2}{g} \frac{\partial g}{\partial\phi}
\vec{E}\cdot \vec{\nabla}\phi,
\label{as3}
\end{eqnarray}
where $\vec{v}$ is the bulk velocity of the plasma.
The spatial gradients used in Eqs. (\ref{as1})--(\ref{as3}) are
defined according to the metric (\ref{metric}). In Eq. (\ref{s1}) 
the quantity ${\cal H} = \partial{\ln{ a}}/\partial\eta$ has also been introduced.
${\cal H}$ is the Hubble factor in conformal time which is 
related to the Hubble factor in cosmic time as $ {\cal H} =  H/a$ where 
$H= \partial{\ln{ a}}/\partial t$.

Once the background geometry is specified we are interested in the 
situation when the gauge field background is vanishing and the only 
fluctuations are  the ones associate with the vacuum state of the 
Abelian gauge fields. Hence,  Eqs. (\ref{as1})--(\ref{as3}) 
allow to compute the evolution of $\phi$ and the 
associated evolution of the two-point function  of the gauge field 
fluctuations. 

Suppose that $\phi$  is originally displaced from the minimum 
of its potential. As far as the zero mode of $\phi$ is concerned the 
system of equations can be further simplified: 
\begin{eqnarray}
&& \frac{\partial^2\vf}{\partial\eta^2} 
+  2 {\cal H} \frac{\partial \vf}{\partial\eta}  
+ \frac{\partial V}{\partial \phi} a^2  
=  \frac{ 1}{ 2 g^3 a^2 }
\frac{\partial g}{\partial \phi} \biggl[ \vec{E}^2 - \vec{B}^2 \biggr],
\label{s1}\\
&& \frac{\partial{\vec{B}}}{\partial\eta} = -\vec{\nabla}
\times \vec{{E}}, 
\label{s1b}\\
&&\frac{\partial}{\partial\eta}\biggl[ \frac{1}{g^2(\vf)} \vec{E}\biggr] 
 =
\frac{1}{g^2(\vf)} {\vec{\nabla}}\times \vec{B}, 
\label{s2}\\
&& {\vec{\nabla}}\cdot{\vec{B}}=0,\,\,\,\,\,
{\vec{\nabla}}\cdot {\vec{E}}=0,
\label{s3}
\end{eqnarray}
By now combining together the modified 
Maxwell's equations we obtain the evolution 
of the magnetic fields
\begin{equation}
\vec{B}'' - 2 \frac{g'}{g} \vec{B}' - \nabla^2 \vec{B} = 0,
\end{equation}
where the prime denoted derivation with respect to the conformal time 
coordinate (the over-dot will denote, instead, derivation with respect 
to cosmic time).

Once the evolution of the metric is specified, Eq. (\ref{s1}) 
dictates a specific evolution for $\phi$ and the evolution 
of $\phi$ will determine, in its turn, the evolution 
of the gauge fields. The interesting initial conditions 
for the system are the ones where 
the classical gauge field background vanishes. Thus,
when the homogeneous component of $\phi$ starts its evolution during 
the de Sitter phase, quantum mechanical 
fluctuations will be postulated as 
initial conditions of gauge inhomogeneities. 

When the background geometry 
evolves from the de Sitter phase to the subsequent 
epoch, massive quanta of $\phi$ are produced. The amount of the 
produced inhomogeneous modes of $\phi$ can be computed and it will be shown 
that the associated energy density will always be smaller than the 
one of the homogeneous mode. 
This analysis will be one of the subjects 
discussed in Section III.

\renewcommand{\theequation}{3.\arabic{equation}}
\setcounter{equation}{0}
\section{Constraints on the evolution of the gauge coupling} 
Sub-millimiter 
tests of the Newton's law show  that no deviations are 
observed  down to
distances as small as $0.1$ mm \cite{mmm}. Therefore, $\phi$ 
should be massive. 
Of course the potential of $\phi$ may be much more complicated than the one 
provided by a simple mass term. However, for sake of 
simplicity, a massive scalar will be analyzed since, already in this case, 
interesting effects can be analyzed. In spite of the fact that this choice 
is apparently simple, various constraints on the scalar mass appear.

\subsection{Evolution of the homogeneous mode}
Suppose  that
the potential term driving the evolution 
of the gauge coupling is simply
\begin{equation}
 V(\vf) \sim \frac{m^2}{2} \vf^2.
\end{equation}
During the inflationary stage of expansion the scale factor evolves 
as $a(\eta) = (- \eta_1/\eta)$ for $\eta < -\eta_1$.
The evolution of $\vf$ is obtained by solving 
\begin{equation}
\vf''  + \frac{2}{\eta} \vf' + \frac{\mu^2}{\eta^2} \vf  =0,
\label{inf}
\end{equation}
where $\mu = m/H_1$ and where the relation ${\cal H} \sim \eta^{-1} \sim 
a H$ has been used. If $\mu \ll 1 $,  for $\eta < - \eta_1$
the solution of Eq. (\ref{inf}) can be written as 
\begin{equation}
 \vf_{\rm i}(\eta) = \vf_1- \vf_2\biggl(-\frac{\eta}{\eta_1}\biggr)^3.
\label{sol1}
\end{equation}
The end of the inflationary stage of expansion 
may not be  directly followed by the radiation dominated phase. 
In the intermediate  phase the scalar mass is still small than the 
curvature scale 
but the curvature decreases, in general,  faster than during 
the inflationary phase since the 
background is neither of de Sitter nor of quasi-de Sitter type.
The evolution of the scale factor 
can be parametrized as $ a(\eta) \sim \eta ^{\alpha}$ where, in order 
to fix the ideas, $\alpha \sim 2$ could be assumed \footnote{The case 
$\alpha =2$ corresponds to a matter-dominated intermediate stage.}. 
The 
evolution of $\phi$ will simply be 
\begin{equation}
\phi_{\rm rh}(\eta) = \phi_1 + \phi_2 \biggl[ 
\frac{ 2 \alpha - 4}{2 \alpha - 1} + \frac{ 3}{ 2\alpha -1} \biggl( 
\frac{\eta }{\eta_1}\biggr)^{2 \alpha -1} \biggr],\,\,\,\,\, \eta_{1} <
\eta <\eta_{\rm r},
\label{sol2}
\end{equation}
where the continuity between Eq. (\ref{sol1}) and Eq. (\ref{sol2}) has been 
required, so that $\phi_{\rm i} ( -\eta_1) = \phi_{\rm rh}(\eta_1)$ and 
 $\phi'_{\rm i} ( -\eta_1) = \phi'_{\rm rh}(\eta_1)$. 

After $\eta_{\rm r}$
the background enters a radiation dominated phase and 
the evolution of $\vf$ can be 
explicitly solved in cosmic  time. The equation 
for $\vf$ , in this phase, is given by
\begin{equation}
\ddot{\vf} + 3 H \dot{\vf} + m^2 \vf  =0,\,\,\,\,\, H= \frac{\dot{a}}{a}, 
\label{rm1}
\end{equation}
which in terms $ \Phi = a^{\frac{3}{2}} \vf $, becomes 
\begin{equation}
\ddot{\Phi} + \biggl[m^2 - \frac{3}{2} \dot{H} - \frac{9}{4} H^2 \biggr] 
\Phi=0.
\label{rm2}
\end{equation}
In the radiation-dominated stage of expansion 
Eq. (\ref{rm2}) becomes 
\begin{equation}
\ddot{\Phi} + \biggl[ m^2 + \frac{3}{16 t^2}\biggr] \Phi =0,
\end{equation}
whose solution can be written in terms of Bessel functions \cite{abr}
\begin{equation}
\Phi(m t) = \sqrt{ m t} \biggl[ A Y_{\frac{1}{4}}( m t) + B 
J_{\frac{1}{4}}( m t) \biggr].
\label{bs}
\end{equation}
For $m t \ll 1$, $\vf$ has a constant mode  and a solution  as 
$ t^{-1/2}$. Recalling 
the relation between cosmic and conformal time and imposing the continuity 
of $\vf$ and $\vf'$ (in $\eta_r$) with the solution of
 Eq. (\ref{sol2}) the following 
form  can be obtained:
\begin{equation}
\vf_{\rm r}(\eta) = \vf_1 + \phi_2 \biggl[ \biggl( 
\frac{2 \alpha -4}{2\alpha -1} + \frac{6 ( 1-\alpha)}{2\alpha -1} \biggl( 
\frac{\eta_1}{\eta_{\rm r}}\biggr)^{2 \alpha -1} \biggr) + 
3 \biggl( \frac{\eta_1}{\eta_{\rm r}}\biggr)^{2\alpha -1} 
\frac{\eta_{\rm r}}{\eta}\biggr],
\label{sol3}
\end{equation}
which is valid for $\eta_{r} < \eta < \eta_{\rm m}$.
The time  $\eta_{\rm m}$ marks the moment where $ H \sim m$.
When $mt > 1$, the regime of coherent oscillations takes over and 
 the solution (\ref{bs}) implies that
 the energy density stored in $\vf$ decreases as $a^{-3}$, meaning 
that $\phi_{\rm c}(\eta) \sim \eta^{-3/2} $. Since the 
coherent oscillations decrease as $a^{-3}$ there will be a typical curvature 
scale $H_{\rm c}$ and a typical time $\eta_{\rm c}$ at which 
the coherent oscillations become dominant with respect to the radiation 
background. This moment is determined by demanding that 
\begin{equation}
H_{\rm r}^2 \,M_{\rm P}^2 \biggl( \frac{a_{\rm r}}{a_{\rm c}}\biggr)^4 \simeq 
m^2 \phi_1^2 \biggl(\frac{a_{\rm m}}{a_{\rm c}}\biggr)^3,
\label{eq1}
\end{equation}
which also implies that 
\begin{equation}
H_{\rm c} \sim m \varphi^4,
\label{hc1}
\end{equation}
where $\varphi = \phi_1/M_{\rm P}$.
Eq. (\ref{hc1}) has been obtained without tuning the 
asymptotic value of $\phi$ to the minimum of its potential. 
If such a  tuning is made, the amplitude of oscillations 
at $\eta_{\rm m}$ will be smaller than $\phi_1$ and it will 
be given, according to Eq. (\ref{sol3}), by 
$\phi_2 (\eta_1/\eta_{\rm r})^{2\alpha -1}$.
Thus, the scale $H_{c}$ will be defined 
by a different relation namely:
\begin{equation}
m^2 \phi_{2}^2 
\biggl(\frac{H_{\rm r}}{H_{1}}\biggr)^{\frac{2(2 \alpha -1)}{\alpha + 1}} 
\biggl(\frac{m}{H_{\rm r}}\biggr) 
\biggl(\frac{ a_{\rm m}}{a_{\rm c}}\biggr)^3 \simeq
H^2_{\rm r} M^2_{\rm P}\biggl(\frac{a_{\rm r}}{a_{\rm c}}\biggr)^4,
\label{eq2}
\end{equation}
leading, ultimately, to
\begin{equation}
H_{\rm c} = m \biggl(\frac{\phi_{2}}{M_{\rm P}}\biggr)^4 
\biggl( \frac{H_{\rm r}}{H_{1}}\biggr)^{\frac{4( 2 \alpha -1)}{\alpha + 1}} 
\biggl( \frac{m}{H_{\rm r}}\biggr)^2.
\label{hc2}
\end{equation}
In the approximation of instantaneous 
reheating [i.e. $\eta_{\rm r} \sim \eta_{1}$],  
$H_{\rm r } \sim H_{1}$. Therefore, from Eq. (\ref{hc2}) $H_{\rm c}$ 
is smaller 
than the value determined in Eq. (\ref{hc1}) by a factor $(m/H_1)^2$.
In the approximation of matter-dominated reheating (i.e. $\alpha \sim 2$),
the result of instantaneous reheating is further suppressed by a 
factor $(H_{\rm r}/H_{1})^4$ as one can easily argue from Eq. (\ref{hc2}).
From Eqs. (\ref{rm1})--(\ref{rm2}), the evolution of $\phi$ will go as 
$\eta^{-3}$ when coherent oscillations start dominating.

In spite of the possible tunings made in the asymptotic values 
of $\phi$, after $\eta_{\rm c}$ there will be a typical time 
at which the field $\phi$ will decay.      
In order not to spoil the light elements 
produced at the epoch of BBN $\vf$ has 
to decay at a scale larger than $H_{\rm ns}\simeq T_{\rm ns}^2/M_{\rm P}$ 
(where $T_{\rm ns} \simeq {\rm MeV}$. Since $\vf$ is only coupled 
gravitationally the typical decay scale will be given by comparing 
the rate with 
the curvature scale giving that 
\begin{equation}
H_{\vf} \sim \Gamma \sim \frac{m^3}{ M_{\rm P}^2} > H_{\rm ns},
\label{d}
\end{equation}
implying that $ m > 10^4$ GeV. This requirement also demands 
that the reheating temperature associated with the decay of $\vf$ 
will be larger than the BBN temperature. 

In order to illustrate some concrete examples of the various 
possibilities implied by our considerations, 
suppose that $m \sim 10^3$ TeV and suppose 
that the asymptotic value of $\vf$ is fine-tuned to its minimum. Furthermore,
suppose that the reheating is instantaneous.
Then, according 
to the picture which has been presented, 
inflation stops at a scale $H_1 \sim 10^{13}$ GeV and 
$\vf$ starts oscillating at a curvature scale $H_{\rm m} \sim 10^3 $ TeV. 
The coherent oscillations will then become dominant at a curvature scale 
$H_{\rm c} \sim 10^{-8}$ GeV (having assumed $\vf_2 \sim M_{\rm P}$). 
The coherent oscillations of $\vf$ will  last down to $H_{\vf} 
\sim 10^{-20} $ 
GeV. After this moment the Universe will be dominated by the radiation
produced in the decay of $\vf$. Notice, for comparison, that the 
BBN curvature scale is $H_{\rm ns}\simeq 10^{-25}$ GeV so 
that the decay occurs 
well before BBN (five orders of magnitude in curvature scale). 

Another illustrative example is the one where $m \sim 10^{6}$ TeV. In this 
case the decay of $\vf$ occurs prior to the EWPT epoch, namely
\begin{equation}
H_{\vf} > H_{\rm ew}.
\label{e}
\end{equation}
In fact 
$H_{\rm ew} = \sqrt{N_{\rm eff}} T^2_{\rm ew }/M_{\rm P} \sim 10^{-17}$ 
GeV (with $N_{\rm eff} = 106.75$ and 
$T_{\rm ew} \sim 100$ GeV) whereas, from Eq. (\ref{d}), $H_{\phi} \sim 
10^{-9} $ GeV. In more general terms we can say that in order to have 
the $\vf$ decay occurring prior to the EWPT epoch 
we have to demand that $H_{\vf}> H_{\rm ew} $ which means 
that $ m > 10^{5}$ TeV. 

In closing this section two general comments are in order.
If no fine-tuning is made in the asymptotic amplitude of $\phi$,
 the typical scale of the coherent oscillations will almost 
coincide with $m$. However, the possibility  
$\varphi\ll 1$  is still left if, for some reason,  we want 
$H_{\rm c} \ll m$.

The decay of $\phi$ and the consequent freezing 
of the gauge coupling should occur prior to the EWPT epoch 
and the baryon number should be generated, in the present context, 
at the electroweak time. 
Suppose, for example, that this is not the case and that 
the BAU has been created prior to the electroweak scale. Suppose, moreover, 
that the decay of $\phi$ occurs after baryogenesis. Then the temperature 
of the radiation gas before the decay of $\phi$ will be 
$T_{\phi} \sim T_{\rm m} (a_{\rm m}/a_{\phi}) \sim m (m/M_{\rm P})^5$. Thus, 
the entropy increase due to the decay of $\phi$ will be $\Delta S \sim (T_{\rm 
decay}/T_{\phi})^3$ where $T_{\rm decay} \sim \sqrt{H_{\phi} M_{\rm P}}$. 
This implies that $\Delta S \sim m/M_{\rm P}$. It has been observed in 
different contexts that in order to preserve a pre-existing BAU one should 
have $\Delta S < 10^{5}$ \cite{l,ts}. Thus, this bound would imply 
$m > 10^{14 }$ GeV. This is the reason why the present analysis will assume 
that the decay of $\phi$ 
occurs prior to the electroweak time and  that the BAU 
is generated at the EWPT or shortly after.
 
\subsection{Evolution of the inhomogeneous modes}
When the Universe passes from the inflationary stage to the 
subsequent radiation dominated expansion, inhomogeneities of the field 
$\phi$ are generated. This may invalidate the original assumptions and 
introduce further complications by adding qualitatively new
constraints on the scenario.

It is useful to recall that
the inhomogeneities of $\phi$ can be interpreted, in the framework of 
second quantization, as quanta of the field $\phi$. Hence, the 
inhomogeneities produced because of the sudden change of the 
geometry  from the de Sitter epoch to the radiation dominated epoch 
can be counted by estimating  the number of quanta produced by the 
sudden change of the geometry according to the well known techniques 
of curved space-times \cite{bir}.

Consider the first order fluctuations of 
the field $\vf$
\begin{equation}
\vf(\vec{x},\eta) = \vf(\eta) + \delta \vf(\vec{x},\eta),
\end{equation}
whose evolution equation is, in Fourier space,
\begin{equation}
\psi'' + 2 {\cal H} \psi' + [ k^2 + m^2 a^2] \psi =0,
\label{fluc}
\end{equation}
where $\psi(k,\eta)$ is the Fourier component 
of $\delta\vf(\vec{x}, \eta)$.
In order to count the number of 
quanta produced during the transition of the 
geometry from the inflationary to the radiation 
dominated stage of expansion the (canonically 
normalized)  amplitude of fluctuations 
$\Psi = \psi a$ should be defined so that  Eq. (\ref{fluc}) becomes:
\begin{equation}
\Psi'' + \bigl[ k^2 + m^2 a^2 - \frac{ a''}{a}\bigr] \Psi =0.
\label{fluc1}
\end{equation}
In the de Sitter stage of expansion Eq. (\ref{fluc1}) reduces to
\begin{equation}
\Psi_{\rm i}'' + \bigl[ k^2 + \frac{\mu^2 - 2 }{\eta^2}  \bigr] \Psi_{\rm i} 
=0,
\label{fluc2}
\end{equation}
whereas during the radiation dominated stage of expansion Eq. (\ref{fluc2})
takes the form 
\begin{equation}
\Psi_{\rm r}'' + \bigl[ k^2 
+ \frac{ \mu^2 (\eta + 2 \eta_1)^2}{\eta_1^4}\bigr] \Psi_{\rm r} =0.
\label{fluc3}
\end{equation}
The solution of Eq. (\ref{fluc2}) 
(with the correct quantum-mechanical normalization for 
$\eta\rightarrow - \infty$)  can be written as 
\begin{equation}
\Psi_{\rm i}(\eta)= \frac{1}{\sqrt{2 k}}\,\, p\,\,\, \sqrt{- x} \,\,\,
H^{(1)}_{\rho}(-x),  
\end{equation}
where $x = k \eta$ and $H^{(1)}_{\nu}$ is the first order Hankel
function \cite{abr}. In the pure de Sitter case, $\rho = 3/2 \sqrt{ 1 -
(4/9) \mu^2}$ and since  $\mu \ll 1$, $\rho \simeq 3/2$; 
$p$ is  a phase factor  which has been chosen in such a way that 
\begin{equation}
p = \sqrt{\frac{\pi}{2}}\,\,e^{i\frac{\pi}{4}( 1 + 2 \rho)}.
\end{equation}
With this choice of $p$ we have that  $\Psi_{\rm i}(\eta) \sim e^{- i
k\eta}/\sqrt{2 k}$ for $\eta \rightarrow - \infty$.

During the radiation dominated stage
 of expansion Eq. (\ref{fluc3}) 
is  the equation of  parabolic cylinder 
functions \cite{abr}. 
The solutions turning into positive and 
negative frequencies for $\eta \rightarrow +\infty$ are then
\begin{eqnarray}
&& f_{\rm r}(\eta) = \frac{ 1}{(2 \gamma )^{1/4} } \, e^{
i\frac{\pi}{8}}  D_{ - i q - \frac{1}{2}}( i e^{- i\frac{\pi}{4}} z),
\nonumber\\
&& f^{\ast}_{\rm r}(\eta) = \frac{ 1}{(2 \gamma )^{1/4} } \, e^{
-i\frac{\pi}{8}}  D_{  i q - \frac{1}{2}}(  e^{- i\frac{\pi}{4}} z),
\end{eqnarray}
where 
\begin{equation}
z = \sqrt{2 \gamma} (\eta + 2 \eta_1),\,\,\,\, q = \frac{k^2}{2
\gamma},
\label{neweq}
\end{equation}
and where $D_{\sigma}$ are the parabolic cylinder 
functions in the Whittaker's notation.
The solution of Eq. (\ref{fluc3}) 
\begin{equation}
\Psi_{\rm r}(\eta) = c_{+}(k) g_{\rm r}(\eta) + c_{-}(k) 
g^{\ast}_{\rm r}(\eta),
\end{equation}
is given in terms of 
 $c_{+}(k)$ and $c_{-}(k)$ which are the two (complex) Bogoliubov 
coefficients satisfying $|c_{+}(k)|^2 - |c_{-}|^2 =1$.
In a second quantized approach $|c_{-}(k)|^2$ is the mean 
number of created quanta, whereas in a semi-classical 
approach $c_{-}(k)$ can be viewed as the coefficient 
parametrising the mixing between positive and negative
frequency modes. In the case $c_{-}(k)\simeq 0$ 
no mixing takes place and no amplification 
is produced. 
In order to determine $c_{\pm}(k)$, $\Psi_{\rm i}(\eta)$ and 
$\Psi_{\rm r}(\eta)$ should be continuously matched 
in $\eta = -\eta_1$, namely 
\begin{eqnarray}
&&\Psi_{\rm i}(-\eta_1) = \Psi_{\rm r}(-\eta_1), 
\nonumber\\
&& \Psi'_{\rm i}(-\eta_1) = \Psi'_{\rm r}(-\eta_1),
\end{eqnarray}
By solving this system, an exact expression  for the
Bogoliubov coefficients is obtained 
which is, in general a function of two 
variables : $\mu= m\eta_1$ and $x_1 = k \eta_{1}$. Since $\mu \ll 1$
the exact result can be expanded, in this limit,
\begin{eqnarray}
&&c_{+}(k) = \pi e^{ i \frac{\pi}{8}} \biggl\{  \frac{i}{ 
 \sqrt{2} \Gamma(\frac{3}{4})} S_2(x_1, \rho) \mu^{-\frac{1}{4}} + 
\frac{ (1 +i)}{2 \Gamma(\frac{1}{4})} [ S_1(x_1, \rho) + S_2(x_1, \rho)] 
\mu^{\frac{1}{4}} \biggr\} + {\cal O}(\mu^{\frac{5}{4}}),
\nonumber\\
&&c_{-}(k) =  \pi e^{ -i \frac{\pi}{8}} \biggl\{ -\frac{ i}{ 
 \sqrt{2} \Gamma(\frac{3}{4})} S_2(x_1, \rho) \mu^{-\frac{1}{4}} + 
\frac{ (i -1)}{2 \Gamma(\frac{1}{4})} [ S_1(x_1, \rho) + S_2(x_1, \rho)] 
\mu^{\frac{1}{4}} \biggr\} + {\cal O}(\mu^{\frac{5}{4}}),
\label{bog1}
\end{eqnarray}
where $S_1(x_1, \rho)$ and $S_2(x_1,\rho) $ contain the explicit 
dependence upon the Hankel's functions:
\begin{eqnarray}
&& S_1(x_1, \rho) = e^{i \frac{\pi}{4} ( 1 + 2 \rho)}
H_{\rho}^{(1)}(x_1),
\nonumber\\
&& S_2(x_1, \rho) = \sqrt{x_1} e^{ i\frac{\pi}{4} ( 1 + 2 \rho)} 
\biggl[ \bigl( \rho + \frac{1}{2}\bigr) \frac{
H_{\rho}^{(1)}(x_1)}{\sqrt{x_1}} - \sqrt{x_1} H^{(1)}_{\rho + 1 } (x_1)
\biggr].
\label{s1s2}
\end{eqnarray}
If $\rho \sim 3/2$ Eq. (\ref{s1s2}) gives
\begin{eqnarray}
&&c_{+}(k) = e^{i \frac{\pi}{8}} \sqrt{\pi}\biggl\{ 
-\frac{ x_1^{ - \frac{3}{2}}}{ 2 \Gamma(\frac{3}{4}) \mu^{ \frac{1}{4}}} 
 + \frac{i\,x_1^{-\frac{1}{2}} }{2 \Gamma( \frac{3}{4}) \mu^{\frac{1}{4}}}
 + \biggl[ \frac{1}{2 \Gamma( \frac{3}{4}) \mu^{\frac{1}{4}}}
+ \frac{( i - 1) \mu^{1/4}}{\sqrt{2} \Gamma(\frac{1}{4})}\biggr]
 \sqrt{x_1}\biggr\}
+ {\cal O}(\mu^{\frac{5}{4}}),
\nonumber\\
&&c_{-}(k) = e^{-i \frac{\pi}{8}} \sqrt{\pi}\biggl\{ 
\frac{x_1^{ - \frac{3}{2}} }{ 2 \Gamma(\frac{3}{4}) \mu^{\frac{1}{4}}} 
- \frac{i\,x_1^{-\frac{1}{2}} }{2 \Gamma( \frac{3}{4}) \mu^{\frac{ 1 }{4}}}
 + \biggl[ 
-\frac{1}{2 \Gamma( \frac{3}{4}) \mu^{\frac{1}{4}}}
+ \frac{( i + 1) \mu^{1/4}}{\sqrt{2} \Gamma(\frac{1}{4})}\biggr]
 \sqrt{x_1}\biggr\}
+ {\cal O}(\mu^{\frac{5}{4}}).
\label{bogds}
\end{eqnarray}
In the limit $ x_1 = k\eta_1 \ll 1$ the mean number of 
created quanta can be finally approximated as 
\begin{equation}
\overline{n}(k) \simeq |c_{-}(k)|^2= q |k \eta_1|^{-2 \rho} \mu^{-1/2} 
\end{equation}
where $q$ is a numerical coefficient of the 
order of $10^{-2}$. 
The energy density of the created (massive) quanta 
can be estimated from 
\begin{equation}
d\rho_{\psi} = \frac{d^3 \omega}{(2\pi)^3}~ m~ \overline{n}(k)
\end{equation}
where $ \omega = k/a$  is the physical momentum. 
In the case of a de Sitter phase ($\rho = 3/2$) the typical 
energy density of the produced fluctuations 
is 
\begin{equation}
\rho_{\psi}(\eta) \simeq 
q \,m\, H_1^3\, \biggl( \frac{m}{H_1}\biggr)^{-1/2} \biggl(\frac{a_1}{a}
\biggr)^3
\end{equation}
The produced massive quanta may become dominant. If they become dominant 
after $\vf$ already decayed they will not lead to further 
constraints on the scenario. If they become dominant prior to the 
decay of $\vf$ further constraints may be envisaged. 
The scale at which the massive fluctuations become 
dominant with  respect to the radiation background can be determined 
by requiring that $\rho_{\psi}(\eta_{\ast}) \simeq \rho_{\gamma} 
(\eta_{\ast})$ implying that 
\begin{equation}
q\,m \,H_1^3 \biggl(\frac{m}{H_1}\biggr)^{-1/2} 
\biggl(\frac{a_1}{a_{\ast}}\biggr)^3 
\simeq H_1^2 \, M_{\rm P}^2 \biggl(\frac{a_1}{a_{\ast}}\biggr)^4,
\end{equation}
which translates into
\begin{equation}
H_{\ast}  \simeq q^2  \, m \epsilon^4,
\label{inmo}
\end{equation}
where $\epsilon = H_1/M_{\rm P}$.
In order to make sure that the non-relativistic modes 
will become dominant after $\vf$ already decayed 
 $H_{\ast} < H_{\vf}$ should be imposed, that is to say 
$m > 10^{2}$ TeV for $\epsilon\sim 10^{-6}$.

The maximum tolerable amount of entropy, in order not to wash-out any 
preexisting BAU is  model-dependent 
but, in general, $\Delta S < 10^{5}$ seems to be acceptable \cite{r,l,ts}.
Defining $T_{\vf}$ as the radiation gas already present at the 
scale $H_{\vf}$, the entropy increase from $T_{\vf}$ to 
$T_{\rm decay} \simeq\sqrt{H_{\vf} M_{\rm P}}$ is of the order of
\begin{equation}
\Delta S = \biggl(\frac{T_{\rm decay}}{T_{\vf}}\biggr)^3, 
\end{equation}
where 
\begin{equation}
T_{\vf} =T_{\ast} \biggl(\frac{a_{\ast}}{a_{\vf}}\biggr)\simeq m~\xi^{1/6} 
~\epsilon^{-1/2},
\end{equation}
where $\xi = m/M_{\rm P}$.
Demanding that $\Delta S < 10^{5}$ implies that 
\begin{equation}
\xi > 10^{-10} \epsilon^3. 
\label{en}
\end{equation}
Taking, as usual, $\epsilon \simeq 10^{-6}$,  $m > 10 $ GeV.

Hence, if the constraints pertaining 
to the homogeneous mode are enforced, the bounds coming from the 
inhomogeneous modes do not invalidate the conclusions of the analysis. 
According to the logic expressed in Eq. (\ref{e}), an 
 illustrative  example  is the case where 
$m> 10 ^{5} $ TeV and the BAU is generated after EWPT. In this case
the bounds obtained in the present section are satisfied and  the analysis 
of the evolution of the inhomogeneous modes shows that the qualitatively 
new bounds introduced in the picture are less constraining than 
the ones obtained in 
the analysis of the dynamics of the homogeneous mode. 

\renewcommand{\theequation}{4.\arabic{equation}}
\setcounter{equation}{0}
\section{Evolution of the gauge field fluctuations}
The  evolution of the field $\phi$ during and after 
the de Sitter stage implies, according to Eqs. (\ref{sol1})--(\ref{sol3}), 
that the two-point  function of the gauge field fluctuations may very well 
grow. 

For $\eta < -\eta_1$, a rough approximation 
suggests that  the effect of the ohmic current is not present 
and  the gauge field is in the vacuum state with 
$k/2$ energy in each of its modes. By promoting the classical
fields to quantum mechanical operators we have that 
the physical polarizations of the magnetic field 
can be written as 
\begin{equation}
\hat{b}_{i}(\vec{x}, \eta) = \int \frac{d^3 k}{(2 \pi)^{3/2}}  \sum_{\alpha} 
e^{\alpha}_{i}\bigl[ \hat{a}_{k,\alpha} 
b(k\eta) e^{i 
\vec{k}\cdot\vec{x}} + \hat{a}_{-k,\alpha}^{\dagger} b^{\ast}(k\eta) e^{- 
i \vec{k}\cdot\vec{x}}\bigr],
\label{ex1}
\end{equation}
where $b(k\eta) = B(k\eta)/g(\eta)$ obey the equation
\begin{equation}
b'' + \bigg[ k^2 - 2\biggl(\frac{g'}{g}\biggr)^2 + \frac{g''}{g} \biggr] b=0.
\label{b}
\end{equation}
Notice that $b(k\eta)$  are the correct 
normal modes whose limit (for $\eta\rightarrow -\infty$) 
should be normalized to $\sqrt{k/2} e^{- i k\eta}$. 
The two-point correlation function of the magnetic fluctuations
can then be expressed as 
\begin{equation}
{\cal G}_{i j} (\vec{r},\eta) \equiv
\langle \hat{b}_{i}(\vec{x}, \eta)\hat{b}_{j}(\vec{x}+ \vec{r}, \eta)\rangle= 
\int \frac{d^3 k}{(2 \pi)^3} P_{i j}(k) 
b(k,\eta) b^{\ast}(k,\eta) e^{i \vec{k} \cdot \vec{r}},
\label{tp}
\end{equation}
where 
\begin{equation}
P_{i j}(k) =  \bigl( \delta_{i j} - \frac{ k_i k_j}{k^2}\bigr).
\end{equation}
The magnetic energy density, derived 
from the energy-momentum tensor corresponding to the action of 
Eq. (\ref{action1}), is related 
to the trace  of the correlation function reported in Eq. (\ref{ex1}) over 
the physical polarizations:
\begin{equation}
\rho_{B}(r,\eta) = \int \rho_{B}(k, \eta) \frac{ \sin{kr}}{k r} \frac{ dk}{k}
\end{equation}
where 
\begin{equation}
\rho_{B}(k,\eta) =\frac{1}{ \pi^2} k^3 |b(k,\eta)|^2. 
\label{endens}
\end{equation}
A necessary condition in order to assess that gauge field fluctuations 
grow during the de Sitter phase is that the two-point function increases 
in the limit $\eta \rightarrow - \eta_1$ \cite{msmax0,msmax}.

Suppose that the gauge coupling decreases with monotonic dependence 
upon the field $\phi$, namely
\begin{equation}
g(\eta) = \bigl(\frac{\vf - \vf_1}{M_{\rm P}}\bigr)^{\frac{\lambda}{2}},\,
\,\,\lambda > 0.
\label{g1}
\end{equation}
This parametrisation is purely phenomenological, however, it allows 
to take into account, at once, some physically interesting 
cases like the one suggested by the low-energy string effective action where, 
in the limit of $\vf/M_{\rm P} <1$, $g^2(\phi) \sim \phi$.

Using Eq. (\ref{sol1}) and Eq. (\ref{g1}) into Eq. (\ref{b}) the time 
evolution of the normal modes of the magnetic field can be found analytically 
since the specific form of Eq. (\ref{b}) falls in the same category 
of Eq. (\ref{fluc2}). Hence, 
\begin{equation}
b(k, \eta) = N \sqrt{k \eta} H^{(2)}_{\nu}(k\eta),\,\,\, N= 
\frac{\sqrt{k\pi}}{2} 
e^{- i\frac{\pi}{4} ( 1 + 2 \nu)},
\label{solbb}
\end{equation}
with $\nu = (3\lambda +1)/2$ and where $H_{\nu}^{(2)}(k\eta)$ the Hankel 
function of second kind \cite{abr}. Notice that the normalization $N$ has been 
chosen in such a way that for $\eta \rightarrow -\infty$ the correct 
quantum mechanical normalization is reproduced.
Consequently, following Eq. (\ref{tp}), the two-point function evolves as 
\begin{equation}
\lim_{\eta \rightarrow -\eta_{1} } {\cal G}_{i j}(r, \eta) 
\sim 
\biggl|\frac{\eta}{\eta_{1}}\biggr|^{- 3 \lambda}.
\end{equation}
Since the two-point function increases magnetic fluctuations 
are generated. 

For sake of completeness  the case of increasing 
 gauge coupling will now be examined using the following phenomenological 
parameterisation
\begin{equation}
g(\eta) = \bigl(\frac{\vf - \vf_1}{M_{\rm P}}\bigr)^{-\frac{\delta}{2}},\,
\,\,\delta > 0.
\label{g2}
\end{equation}
Again different scenarios can be imagined. For instance, one coould 
argue in favour of scenarios where the gauge coupling depends upon $\phi$ 
(or upon $\eta$)  
in a highly non-monotonic way. For the illustrative 
purposes of the present investigation it is however sufficient 
to focus the attention on the case of monotonic dependence.

Following now the same steps outlined in the case of decreasing gauge 
coupling the evolution of the two-point function can be obtained
\begin{equation}
\lim_{\eta \rightarrow -\eta_{1} } {\cal G}_{i j}(r, \eta) 
\sim \biggl|\frac{\eta}{\eta_{1}}\biggr|^{2- 3 \delta}.
\label{delone}
\end{equation}
for $\delta > 1/3$ and 
\begin{equation}
\lim_{\eta \rightarrow -\eta_{1} } {\cal G}_{i j}(r, \eta) 
\sim \biggl|\frac{\eta}{\eta_{1}}\biggr|^{ 3 \delta}.
\label{deltwo}
\end{equation}
for $\delta < 1/3$. If $\delta < 1/3$ the correlation 
function decreases and this signals that large scale magnetic fields 
are not produced. 

The back-reaction of the produced fluctuations can be 
safely neglected in de Sitter space. Looking at Eqs. (\ref{Eq1}) and 
(\ref{s1}) it can happen that if the magnetic fluctuations 
 grow too much the term at the right hand side will become of the 
same order of the others. This is not the case. Using 
the conventions of this Section together with the explicit 
form of the scale factor in the de Sitter phase it can be 
shown that
\begin{equation}
\frac{1}{g^3 a^2} \frac{\partial g}{\partial \phi} \vec{B}^2 \sim 
\biggl|\frac{\eta}{\eta_1} \biggr|^{-2\nu} |k\eta_1|^{5 - 2 \nu}
\label{back}
\end{equation}
where $\nu$ is determined from the specific power dependence 
of the coupling as a function of $\phi$. Since we are interested 
in large scale modes we have $k\eta_1 \ll 1$. Therefore the back reaction 
effects are relevant towards the end of the de Sitter phase (i.e. $\eta \sim 
-\eta_1$ ) and for $k \sim \eta_1^{-1}$, namely exactly 
for the modes not relevant for the present investigation.

After the end of inflation the onset of the conductivity
dominated regime may not be instantaneous. In this 
case after $\eta_1$ the presence of a reheating 
phase should be taken into account. Suppose, for instance, that 
$g(\eta)$ decreases according to Eq. (\ref{g1}). Suppose, moreover, that 
during reheating the background is dominated by the coherent 
oscillations of the inflaton. In this case the effective evolution 
of the geometry will be dominated by matter with $a(\eta) \sim \eta^2$.
According to Eq. (\ref{sol2}) (with $\alpha \sim 2$), $\phi \sim \eta^{-3}$.
 The value of the magnetic 
inhomogeneities at $\eta_{\rm r}$ will be given by solving 
Eq. (\ref{b})  
\begin{equation}
b(k,\eta_{\rm r}) \sim A_1 
\biggl(\frac{\eta_{\rm r}}{\eta_1}\biggr)^{\frac{3}{2}\lambda} + 
A_2  \biggl(\frac{\eta_{\rm r}}{\eta_1}\biggr)^{1 - \frac{3}{2}\lambda},
\label{rhb}
\end{equation}
where $A_1$ and $A_2$ are two arbitrary constants. Depending upon the value 
of $\lambda$ the fastest growing solution is selected in Eq. (\ref{rhb}).
This phase may induce further amplification on the two-point function since 
its main effect is to delay the conductivity-dominated regime.

\subsection{Generalized MHD equations}

For $\eta > \eta_{r}$ the role of the Ohmic diffusion becomes 
important and the evolution of the 
magnetic inhomogeneities will be described by the MHD equations generalized
 to the case of time varying gauge coupling.

The ordinary (i.e. fixed coupling) MHD treatment is an effective 
description valid for length scales larger than the Debye radius
and for frequencies smaller than the iono-acoustic frequency.
This means that 
MHD is accurate in reproducing the spectrum of plasma 
excitations obtained from the full kinetic (Vlasov-Landau) 
approach but only for sufficiently low frequencies and for sufficiently
 large scales. 

Implicit in the ordinary MHD analysis is the assumption 
that the plasma has to be  electrically neutral
(${\vec{\nabla}}\cdot {\vec{E}}=0 $) over  length scales 
larger than the Debye radius. Thus, this system of equations 
cannot be applied for distances shorter than the Debye radius and for 
frequencies larger than the plasma frequency  where a kinetic 
description should be employed. 

MHD equations can be derived from a microscopic (kinetic) 
approach and also from a macroscopic approach where 
the displacement current is neglected \cite{k}. 
If the displacement 
current is neglected the electric field can be expressed 
using the Ohm law and the magnetic diffusivity equation can be derived
\begin{equation}
\frac{\partial \vec{B}}{\partial\eta} = \vec{\nabla} \times(\vec{v} 
\times \vec{B}) + \frac{1}{\sigma} 
\nabla^2 \vec{B}.
\label{mdiff}
\end{equation}
The term containing the bulk velocity field is called dynamo term 
and it receives contribution provided parity is globally broken over the 
physical size of the plasma. In Eq. (\ref{mdiff}) 
the contribution 
containing the conductivity is usually called magnetic diffusivity 
term.

In the superconducting (or ideal) approximation the 
resistivity of the plasma goes to zero and the induced (Ohmic)
electric field is  orthogonal both to the bulk velocity of the plasma and 
to the magnetic field [i. e.  $\vec{E} \simeq - \vec{v}\times \vec{B}$]. 
In the real (or resistive) approximation the resistivity may be very small 
but it is always finite and  the Ohmic field can be expressed as 
\begin{equation}
\vec{E} \simeq \frac{\vec{\nabla}\times \vec{B}}{\sigma}
- \vec{v}\times \vec{B}.
\end{equation}

If the gauge coupling changes with time the system of equations obtained by 
neglecting the displacement current receives new contributions and the 
relevant equations can be obtained, in the resistive approximation: 
\begin{eqnarray}
&& \vec{\nabla}\times \vec{E} = - \frac{\partial \vec{B}}{\partial\eta},
\label{mh1}\\
&& \vec{ E} = \frac{\vec{J}}{\sigma} - \vec{v} \times \vec{B},
\label{mh2}\\
&& \frac{1}{g^2} \vec{\nabla}\times \vec{B}= \vec{J} - 2 \frac{g'}{2 g^3} 
\vec{E},
\label{mh3}
\end{eqnarray} 
Using Eqs. (\ref{mh1})--(\ref{mh3})  the generalized magnetic diffusivity 
equation can be obtained:
\begin{equation}
\bigl( 1 - \frac{2}{\sigma~g^2} \frac{g'}{g}\bigr) 
\frac{\partial \vec{B}}{\partial\eta} = \vec{\nabla} \times(\vec{v} 
\times \vec{B}) 
+ \frac{1}{\sigma g^2} \nabla^2 \vec{B}.
\label{modMHD}
\end{equation}
Notice that Eq. (\ref{modMHD}) 
reproduces  Eq. (\ref{mdiff}) if  $g'\rightarrow 0$.

Suppose now that the plasma, whose effective Ohmic description has been 
presented, is relativistic. In the case when the coupling  is 
constant  $\sigma$ is constant and 
it is given by 
\begin{equation}
\sigma \equiv \sigma_{c}(\eta) a(\eta),
\end{equation}
where $\sigma_{c} \sim T/g^2$ scales as the inverse of $a(\eta)$ if the 
evolution of the Universe is, to a good approximation, adiabatic.

If $g$ is not constant, $\sigma$ is not constant 
anymore but it decreases if the gauge coupling increases and, vice versa,
it increases if the gauge coupling decreases. 
In spite of this, in the generalized MHD equations, the combination 
which appears is always $\sigma g^2$ which is roughly constant 
for an adiabatically expanding Universe.

In the approximation 
of instantaneous reheating, 
 the solution of Eq. (\ref{modMHD}) is given by
\begin{equation}
B(k,\eta) = B(k,\eta_1) e^{ -\int \frac{k^2}{\sigma g^2 - 
2 \frac{g'}{g}} d\eta}.
\label{solmhd}
\end{equation}
According to Eqs. (\ref{hc2})--(\ref{d}), in order to get the coupling 
frozen prior to $H_{\rm ew} \sim 10^{-17} $ GeV, $m > 10^{5} $ TeV shall
be required. 
If the gauge coupling is always decreasing
as a function of $\eta$, it can be  parametrized by Eq. (\ref{g1}). 
Hence Eq. (\ref{solmhd}) can be evaluated by using the 
explicit evolution of $\phi$ as obtained from 
 Eqs. (\ref{bs})--(\ref{sol3}) implying that 
$\phi_{\rm r} \sim \eta^{-1}$, $\phi_{\rm m} \sim \eta^{-3/2}$ and 
$\phi_{\rm c} \sim \eta^{-3}$. The result is 
\begin{equation}
B(k,\eta_0) = {\cal I}(\eta_{1}, \eta_{\rm m}, \eta_{\vf}, \eta_{\rm c}) 
e^{ - \frac{k^2}{\sigma g^2}(\eta_1 + \eta_0)}
\label{soleta0}
\end{equation}
where $\eta_0$ is the present time and 
\begin{equation}
{\cal I}(\eta_{1}, \eta_{\rm m}, \eta_{\vf}, \eta_{\rm c}) = 
\bigl[\bigl( \frac{ \lambda + \sigma g^2 \eta_1}{\lambda 
+\sigma g^2 \eta_{\rm m}}\bigr)
\bigl( \frac{ 3\lambda + 2 \sigma g^2 \eta_{\rm m}}{3\lambda 
+2 \sigma g^2 \eta_{\rm c}}\bigr)^{\frac{3}{2}}
\bigl( \frac{ 3\lambda +  \sigma g^2 \eta_{\rm c}}{3\lambda 
+ \sigma g^2 \eta_{\vf}}\bigr)^{3}
 \bigr]^{- \lambda \bigl[\frac{k}{\sigma g^2}\bigr]^2}.
\label{calI}
\end{equation}
Concerning  Eqs. (\ref{soleta0})--(\ref{calI}) few comments are in order.
From Eq. (\ref{soleta0}) all the modes 
\begin{equation}
k^2 >k^2_{\sigma} \sim \frac{\sigma g^2}{\eta_0} 
\end{equation}
are suppressed by the effect of the conductivity. The 
present value of $\omega_{\sigma}(\eta_0)$ 
\footnote{ With $\omega(\eta) \sim k/a(\eta)$ the 
physical momentum will be denoted.} can be estimated by recalling 
that $1/\eta_0 \sim H_0 a_0$ where $H_0\sim 10^{-61}\, M_{\rm P}$.
Thus $\omega_{\sigma} \sim 10^{-3} $ Hz. Present modes of 
the magnetic fields are dissipated if $\omega > \omega_{\sigma}$. 

As far as the problem of galactic magnetic fields is concerned, 
 the relevant set of scales  range around the Mpc corresponding 
to present modes of the magnetic field $\omega_{\rm G} \sim 10^{-14} $ Hz, 
i.e. $\omega_{\rm G} \ll \omega_{\sigma}$. 
 
\renewcommand{\theequation}{5.\arabic{equation}}
\setcounter{equation}{0}
\section{Estimates of large scale magnetic fields}
In the approximation of instantaneous reheating
the typical (present) frequency corresponding to the 
end of inflation can be computed and it turns out to be
\begin{equation}
\omega_1(\eta_0) \sim z_{\rm dec}^{-1}\,\,T_{\rm dec}\,
\,\epsilon^{1/2}\, \xi^{1/3}\varphi^{- \frac{2}{3}}.
\label{om1}
\end{equation}
Since $T_{\rm dec} \sim 0.26$ eV , 
$z_{\rm dec}^{-1}
\,T_{\rm dec} \sim 100 \, {\rm GHz}$. Eq. (\ref{om1}) can be obtained
by red-shifting the highest mode, i.e. $\omega_1(\eta_1) \sim H_1$ 
through the different stages of the evolution of the model, namely, according 
to Eqs. (\ref{hc1}) and (\ref{d}), from $\eta_1$ down to $\eta_{\rm m}$ and 
from $\eta_{\rm m}$ down to $\eta_{rm c}$. Recall that from $\eta_{\rm c}$ 
to $\eta_{\phi}$ the Universe is, effectively, matter dominated.
The other typical frequencies appearing 
in the time evolution of the gauge coupling can be written, in units
of $\omega_1(\eta_0)$, as
\begin{eqnarray}
&& \frac{\omega_{\rm m}(\eta_{0})}{\omega_1(\eta_{0})} =  \epsilon^{-1/2} \,\,\xi^{-1/2}, 
\nonumber\\
&& \frac{\omega_{\rm c}(\eta_{0})}{\omega_1(\eta_0)} = \xi^{1/2}\,\,\epsilon^{-1/2} \,\,
\varphi,   
\nonumber\\
&& \frac{\omega_{\vf}(\eta_0)}{\omega_1(\eta_0)} = \epsilon^{-1/2} \,\,\xi^{7/6}\,\,
\varphi^{2/3},
\label{om}
\end{eqnarray}
where, as in the case of Eq. (\ref{om1}) all the frequencies are evaluated 
at the present time.

In the case of decreasing gauge coupling [described by 
Eq. (\ref{g1})] the amount of generated 
large scale magnetic field can be estimated from Eqs. 
(\ref{b})--(\ref{solbb}) together with Eqs. (\ref{modMHD})--(\ref{soleta0}).
Bearing in mind that the typical frequency scale 
corresponding to $1$ Mpc is $10^{-14}$ Hz we have that the ratio of the magnetic 
to radiation energy density is:
\begin{equation}
r_{B}(\omega_{\rm G}) =f(\lambda)~ \epsilon^{\frac{3}{2} \lambda} \,\,
\xi^{\lambda -\frac{4}{3}} \,\,\varphi^{2\lambda - \frac{8}{3}}
 \,\,10^{-25( 4 -3 \lambda)}\,\,\,
{\cal T}(\omega_{\rm G}),
\label{rb1}
\end{equation}
where 
\begin{equation}
f(\lambda) = \frac{2^{3 \lambda - 1}}{\pi^3} \Gamma^2\biggl(
\frac{3}{2}\lambda + \frac{1}{2} \biggr),
\label{rb2}
\end{equation}
and 
\begin{equation}
{\cal T}(\omega_{\rm G}) \simeq 
e^{ -\frac{\omega^2_{\rm G}}{\omega^2_{\sigma}}}
\bigl[ \bigl(\frac{\omega_{\vf}}{\omega_1} \bigr) 
\bigl( \frac{\omega_{\vf}}{\omega_{\rm m}}\bigr)^{\frac{1}{2}} 
\bigl( \frac{\omega_{\vf}}{\omega_{\rm c}}\bigr)^{\frac{3}{2}} 
\bigr]^{ - \lambda\frac{\omega^2_{\rm G}}{T^2_0}}, 
\label{rb3}
\end{equation}
which means, using Eqs. (\ref{om1})--(\ref{om}), 
\begin{equation}
{\cal T}(\omega_{\rm G}) = e^{ -
\frac{\omega^2_{\rm G}}{\omega^2_{\sigma}}} 
\bigl[ \epsilon^{-1/2} \varphi^{-1} \xi^{5/2}
\bigr]^{ -\lambda\frac{\omega^2_{\rm G}}{T^2_0}}.
\label{rb4}
\end{equation}
Notice that in Eq. (\ref{rb3}) is the present CMB temperature. 

\begin{figure}
\centerline{\epsfxsize = 12 cm  \epsffile{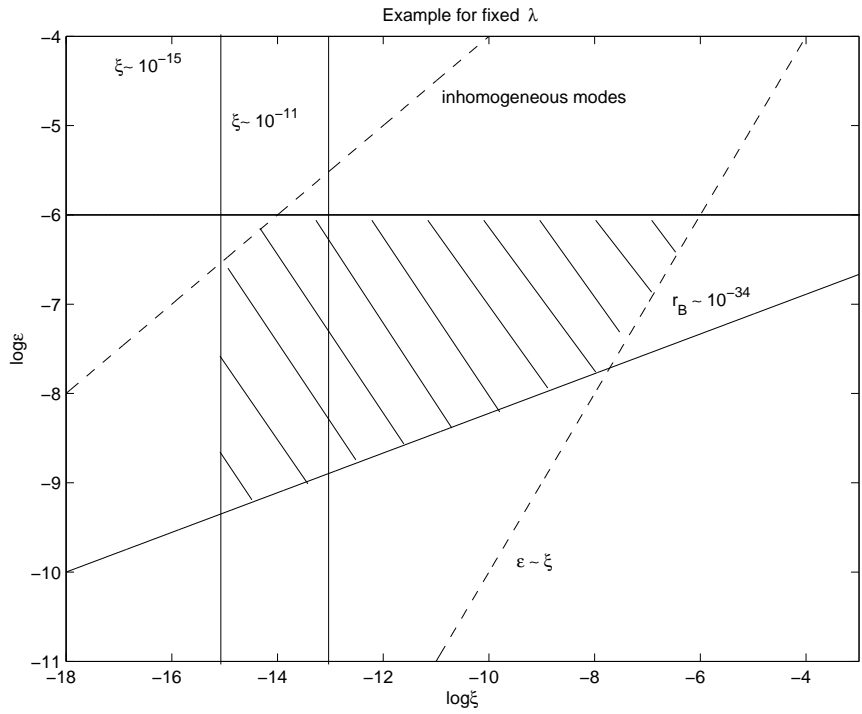}} 
\caption[a]{The shaded area illustrates the region where 
magnetogenesis is possible in the case where $\lambda =1$ and $\varphi \sim 1$.
The vertical lines correspond to the requirements coming from BBN 
and from the EWPT epoch.}
\label{F1}
\end{figure}

\begin{figure}
\centerline{\epsfxsize = 12 cm  \epsffile{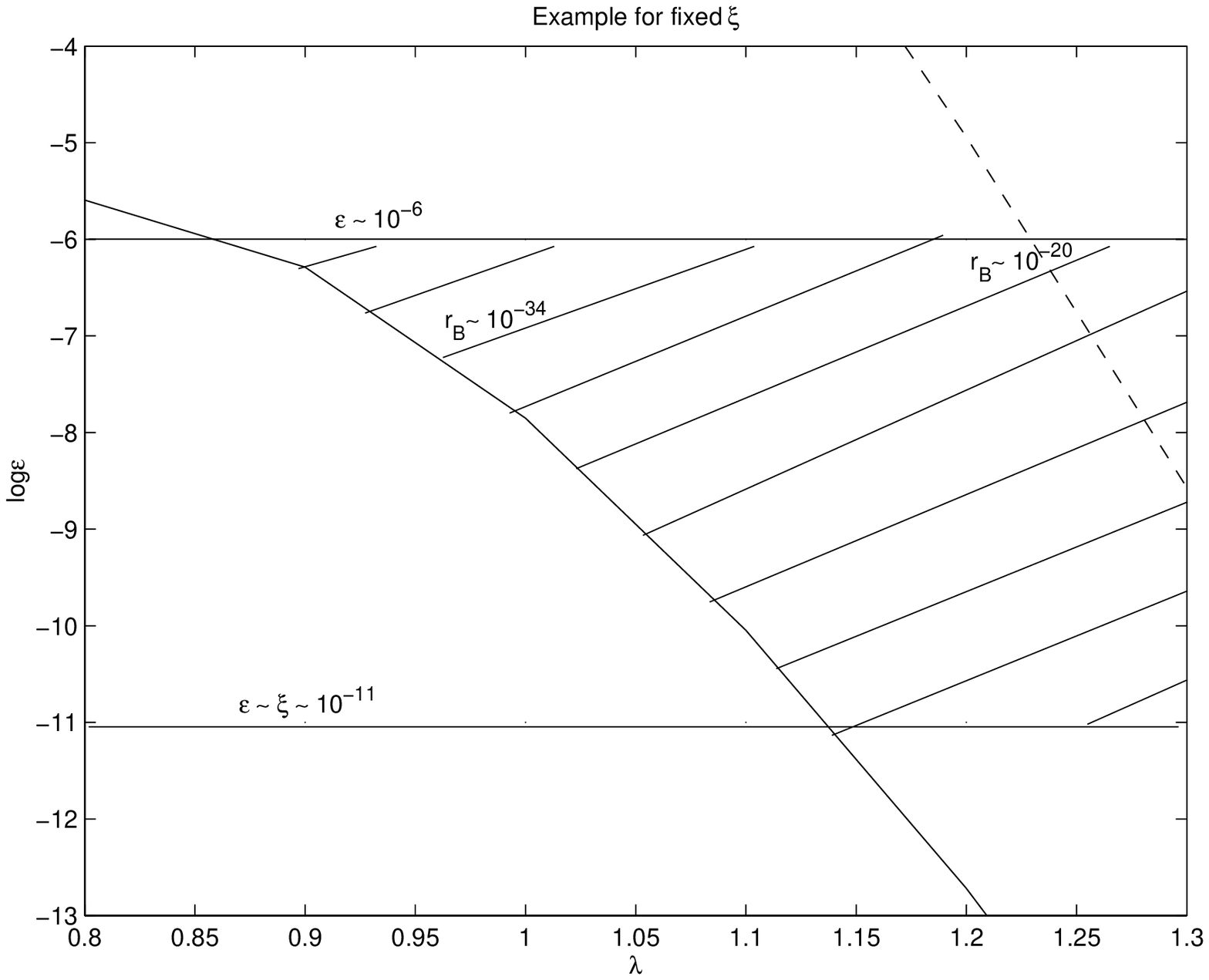}} 
\caption[a]{The region with stripes defines the area 
where  magnetogenesis can occur in the case of fixed mass
[i.e. $\xi \sim 10^{-11}$] and for $\varphi \sim 1$. }
\label{F2}
\end{figure}

In order to illustrate the regions of the parameter space where 
magnetogenesis is possible  $\varphi\sim 1$ will be assumed.
In Fig. \ref{F1} and Fig. \ref{F2} the case of decreasing 
gauge coupling is discussed.
In Fig. \ref{F1}, $\lambda$ is fixed and the exclusion plot is 
given in terms of $\epsilon$ and $\xi$. In particular, for 
illustration, $\lambda=1$ has been chosen. 
The choice of $\lambda\sim 1$ implies that $g^2(\phi) \sim \phi$. Such a  case 
is favoured from the tree-level string effective action where 
the effective coupling can be approximated as $g^2(\phi) 
\sim \phi/M_{\rm P}$ for 
$\phi< M_{\rm P}$. 

The two vertical lines lines mark, respectively, 
the bounds coming from BBN 
[i.e. $\xi > 10^{-15}$, from Eq. (\ref{d})] and from 
the electroweak epoch [i.e. $\xi > 10^{-11}$, from Eq. (\ref{e})]. 
For consistency with the assumptions of Eqs. (\ref{inf})--(\ref{sol1}) 
 $ m /H_1 \ll 1$, implying $\epsilon \gg \xi$. With the lower dashed 
line the bound $\epsilon > \xi$ is reported. Notice that the requirement 
of Eq. (\ref{en}) is not numerically relevant.
The upper 
dashed line is obtained by requiring, according to Eq. (\ref{inmo}),
that $H_{\vf}> H_{\ast}$. Finally, the full (diagonal) line is derived 
by imposing on Eqs. (\ref{rb1})--(\ref{rb4}) the dynamo requirement. 

In Fig. \ref{F2} the value of $\xi$ has been fixed to $10^{-11}$,
as required by the considerations related to the electroweak epoch and 
the magnetogenesis region is described in terms of $\epsilon$ and $\lambda$,
both varying over their physical range.
The two horizontal lines fix the bounds coming from 
$\epsilon \leq 10^{-6}$ and from $\epsilon >\xi$. With the dashed line
the curve $r_{B}(\omega_{\rm G})\sim 10^{-20}$ is denoted. Hence,
values larger than the dynamo requirement are allowed. This 
observation may be relevant in the context of magnetic fields associated with 
clusters. 
In Fig. \ref{F2}, $\lambda$ 
lies in the  range $0< \lambda < 4/3$. This 
choice guarantees the growth of the correlation function 
of the magnetic inhomogeneities during the de Sitter stage.

In order not to conflict with large scale bounds coming from the isotropy 
of the CMB the energy spectra of the produced 
gauge field fluctuations have to decay at large distance scales, implying that 
$0 <\lambda \leq 4/3$. To be compatible with CMB anisotropies 
$r_{\rm B}(\omega_{\rm dec} ) \leq 10^{-10} $ should be imposed (where 
$\omega_{\rm dec} \simeq 10^{-16}$ Hz). If the condition 
$0< \lambda < 4/3$ is enforced, the spectra increase with frequency and the 
possible bounds coming from the anisotropy of the CMB are satisfied.

\begin{figure}
\centerline{\epsfxsize = 12 cm  \epsffile{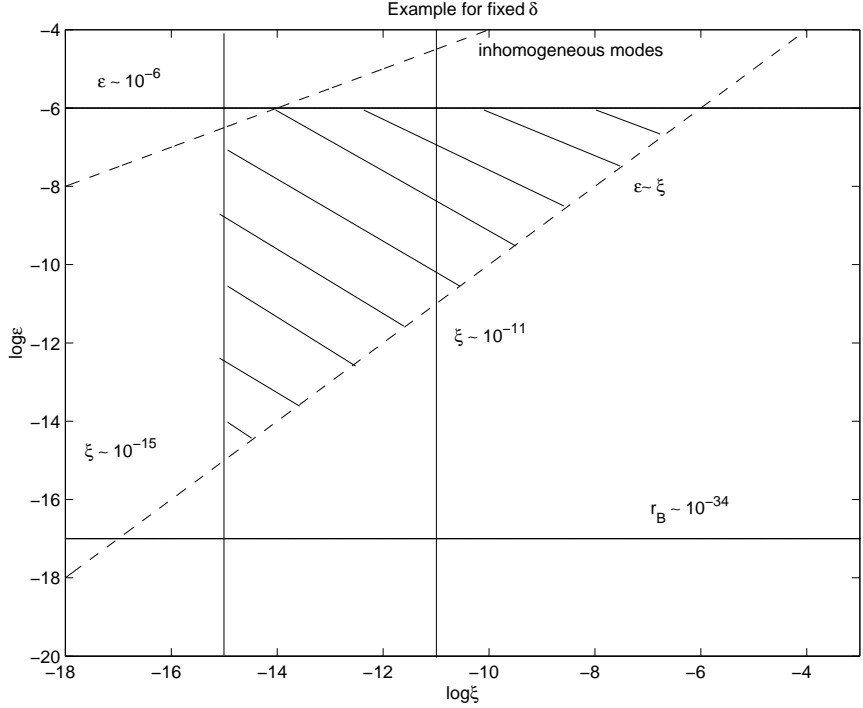}} 
\caption[a]{The magnetogenesis region in the case 
$\delta =2$ (i.e. increasing gauge coupling) 
 and $\varphi\sim 1$. The exclusion plot is then given in terms 
of $\epsilon$ and $\xi$.}
\label{F3}
\end{figure}

\begin{figure}
\centerline{\epsfxsize = 11 cm  \epsffile{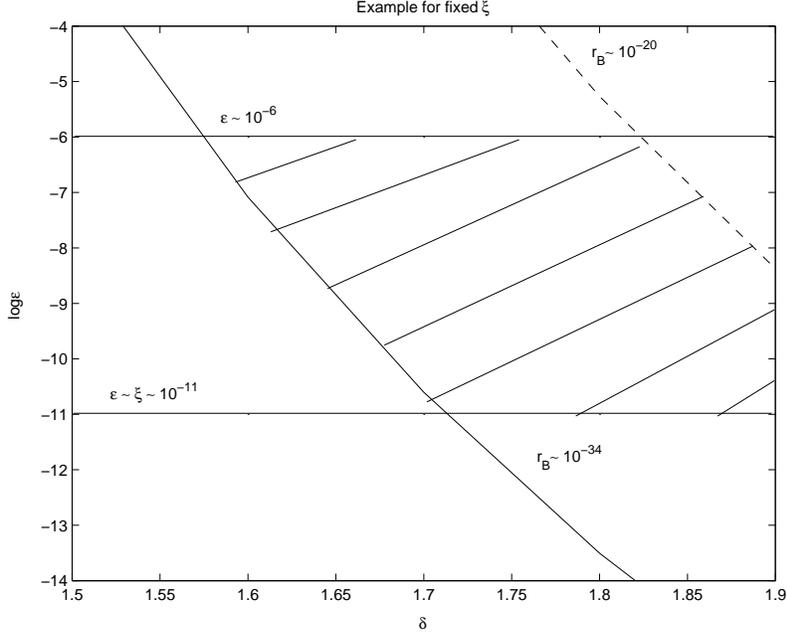}} 
\caption[a]{The magnetogenesis 
region is illustrated  in the case of increasing gauge coupling 
in the ($\epsilon$, $\delta$) plane. Notice that 
in the present example  $\xi\sim 10^{-11}$ (for compatibility with the EW
epoch) and $\varphi \sim 1$.}
\label{F4}
\end{figure}

An analogous estimate can be obtained in the case of 
increasing gauge coupling discussed in Eq. (\ref{g2}). In this case 
\begin{equation}
r_{\rm B} (\omega_{\rm G}) \sim f(\delta) \epsilon^{\frac{3}{2} \delta-1}
\,\,\xi^{\delta -2} \,\,  \varphi^{ 4 - 2 \delta}
\,\,10^{-25 ( 6 - 3 \delta)} {\cal T}(\omega_{\rm G}),
\label{rb5}
\end{equation}
where 
\begin{equation}
f(\delta) = \frac{ 2 ^{3\delta -3}}{\pi^3} \Gamma^2\biggl(\frac{3}{2} 
\delta - \frac{1}{2}\biggr),
\label{rb6}
\end{equation}
and 
\begin{equation}
{\cal T}(\omega_{\rm G}) = e^{ -
\frac{\omega^2_{\rm G}}{\omega^2_{\sigma}}} 
\bigl[ \epsilon^{-1/2} \varphi^{-1} \xi^{5/2}
\bigr]^{\delta\frac{\omega^2_{\rm G}}{T^2_0}}.
\label{rb7}
\end{equation}
In Fig. \ref{F3} and Fig. \ref{F4} the requirements coming from the 
dynamo mechanism as well as the other theoretical constraints are illustrated. For 
both plots $\epsilon < 10^{-6}$ and $\varphi \sim 1$. 
In Fig. \ref{F3} the case 
$\delta =2$ is illustrated. The shaded area selects the region of 
the parameter space where the dynamo requirement  is imposed on 
Eq. (\ref{rb5}). The two  vertical lines 
illustrate the conditions of Eqs. (\ref{d})--(\ref{e}).
As in Fig. \ref{F2} the two dashed lines correspond to the 
constraints coming from the inhomogeneous modes 
and from the condition $\epsilon >\xi$.

The requirement that the spectra decrease at large distance scales 
implies, in this case that $\delta \leq 2$. 
The condition on the growth of the correlation function obtained
 in Eqs. (\ref{delone})--(\ref{deltwo}) imply $\delta > 1/3$. Thus the 
interesting physical range of Eq. (\ref{g2}) and (\ref{rb5}) will be 
$1/3< \delta \leq 2$. 

In Fig. \ref{F4} the parameter space is 
illustrated for fixed values of 
$\xi$, i.e. $\xi\sim 10^{-11}$. 
As in Fig. \ref{F3} the full and dashed curves 
correspond to the dynamo requirements imposed on Eq. (\ref{rb5}).
The shaded area selects the allowed region in the space 
of the parameters where magnetogenesis is possible for 
$10^{-11} <\epsilon< 10^{-6}$. As in the case of Fig. \ref{F2} 
there are regions in the shaded area where values much larger than the 
magnetogenesis requirement are possible (see the dashed line in
Fig. \ref{F4}).

As it has been pointed out in deriving the 
theoretical bounds on the scenario the requirement $\xi \geq 10^{-11}$ 
may be too restrictive since it excludes the variation of the 
gauge coupling at the electroweak time. 
To relax this assumption is possible and it would require 
a precise analysis of the dynamics of the EWPT in the 
presence of time varying gauge coupling. At the moment this 
kind of analysis is not available. 
Summarizing this illustrative discussion,
there are regions 
in the parameter space of the models wher all the theoretical constraints 
are satisfied and where magnetogenesis is possible. In particular, it is 
possible that the gauge coupling freezes prior to the electroweak epoch, 
leading still to magnetogenesis. 

\setcounter{equation}{0}
\section{Concluding remarks}
There are no reasons why the gauge couplings should be constant 
throughout all the history of the Universe. If they are allowed to 
change prior to the formation of the light elements 
they can lead to computable differences in the 
cosmological evolution.

In the present paper the interplay between inflationary 
magnetogenesis and the evolution of the Abelian gauge coupling has been 
addressed. In a phenomenologically reasonable model
of inflationary and post-inflationary evolution
the relaxation of the gauge coupling leads to a growth in the 
correlation function of magnetic inhomogeneities. Large scale 
magnetic fields are then generated. The evolution of the gauge coupling is 
driven, in the present context, by a massive scalar. 

Since the gauge coupling evolves significant changes in the 
evolution of the plasma can be envisaged. In the present 
investigation the ordinary MHD equations have been generalized to the case of 
time evolving gauge coupling but other effects could be envisaged.
In particular, the generlization of the full kinetic approach 
to the case of time evolving ``electron'' charge 
would be of related interest.

The value of large scale magnetic fields produced with this mechanism 
has been estimated. For a broad range of parameters 
the obtained values of the magnetic fields are much larger than the dynamo 
requirements. 

In the present investigation the main assumption has been that
the only gauge coupling free to evolve is the Abelian one. Furthermore,
the parameters of the model have been chosen in such a way that 
the gauge coupling is not dynamical by the onset of the electroweak 
phase transition. It would be interesting to relax both assumptions since 
they may lead to potential differences with the standard scenarios.

Therefore, in many respects, the present investigation is not conclusive. At
the same time it shows that acceptable  models  for the evolution of the 
gauge couplings can be obtained in a standard cosmological framework. 

\section*{Acknowledgements}
The author wishes to thank N. Sanchez and H. de Vega 
for their kind invitation to the Chalonge School and for useful
remarks on the subject of large scale magnetic fields. It is also 
a pleasure to thank interesting conversations with Yoel Raphaeli,
Dan  Boyanovsky, Yuri Pariiski, Sasha  Burinskii and  Peter Coles.

\end{document}